\documentclass{article}
\pdfoutput=1

\usepackage{booktabs} 
\usepackage{array}
\usepackage[a4paper, margin=1in]{geometry}
\usepackage{graphicx}

\usepackage{multirow}
\usepackage[numbers]{natbib}
\usepackage{textcase}
\usepackage{textcomp}

\usepackage[ruled]{algorithm2e} 

\title{Achieving Single-Sensor Complex Activity Recognition from Multi-Sensor Training Data}

\author{Paula Lago$^1$ \and Moe Matsuki$^1$ \and Sozo Inoue$^1$}
\date{%
    $^1$Kyushu Institute of  Technology\\%
}

\begin{document}
\maketitle

\begin{abstract}
In this study, we propose a method for single sensor-based activity recognition, trained with data from multiple sensors. There is no doubt that the performance of complex activity recognition systems increases when we use enough sensors with sufficient quality, however using such rich sensors may not be feasible in real-life situations for various reasons such as user comfort, privacy, battery-preservation, and/or costs. In many cases, only one device such as a smartphone is available, and it is challenging to achieve high accuracy with a single sensor, more so for complex activities. Our method combines representation learning with feature mapping to leverage multiple sensor information made available during training while using a single sensor during testing or in real usage. Our results show that the proposed approach can improve the F1-score of the complex activity recognition by up to 17\% compared to that in training while utilizing the same sensor data in a new user scenario.

\vspace{6pt}

\end{abstract}
\section{Introduction}
Human Activity Recognition (HAR) is the process of automatically identifying what a user is doing from sensor observations. Cameras, wearable sensors, and object-attached sensors have been used for recognizing human activity in different scenarios~\cite{6208895}.
Estimating the activity of a user, correctly enables a wide range of applications, such as, assisting elders with activities in their daily lives~\cite{Hoey2012}, detecting abnormal situations~\cite{ABBATE2012883}, or early disease diagnosis~\cite{Favela2013}.

Traditionally, activity recognition relies on supervised machine learning~\cite{survey-Lara2013}. That is, a set of activity examples, i.e. the \textit{training set}, is used to learn a function to classify new activity examples, i.e. the \textit{test set}.
This model assumes that both the \textit{training set} and the \textit{test set} have the same dimensions; in other words, they share the same number and type of features used to \textit{represent} sensor data. The easiest way to comply with such an assumption is to use the same set of sensors for measuring activities in both the training and test examples. While it is easy to collect training data with multiple sensors, they are rarely used in real-life environments because of the following reasons:
\begin{itemize}
    \item Users may not have multiple devices due to costs or comfort
    \item Users may not want to use some sensors due to privacy concerns
    \item Users may not want to enable multiple sensors due to increased battery consumption
\end{itemize}

\begin{figure*}
    \centering
    \includegraphics[width=0.7\textwidth]{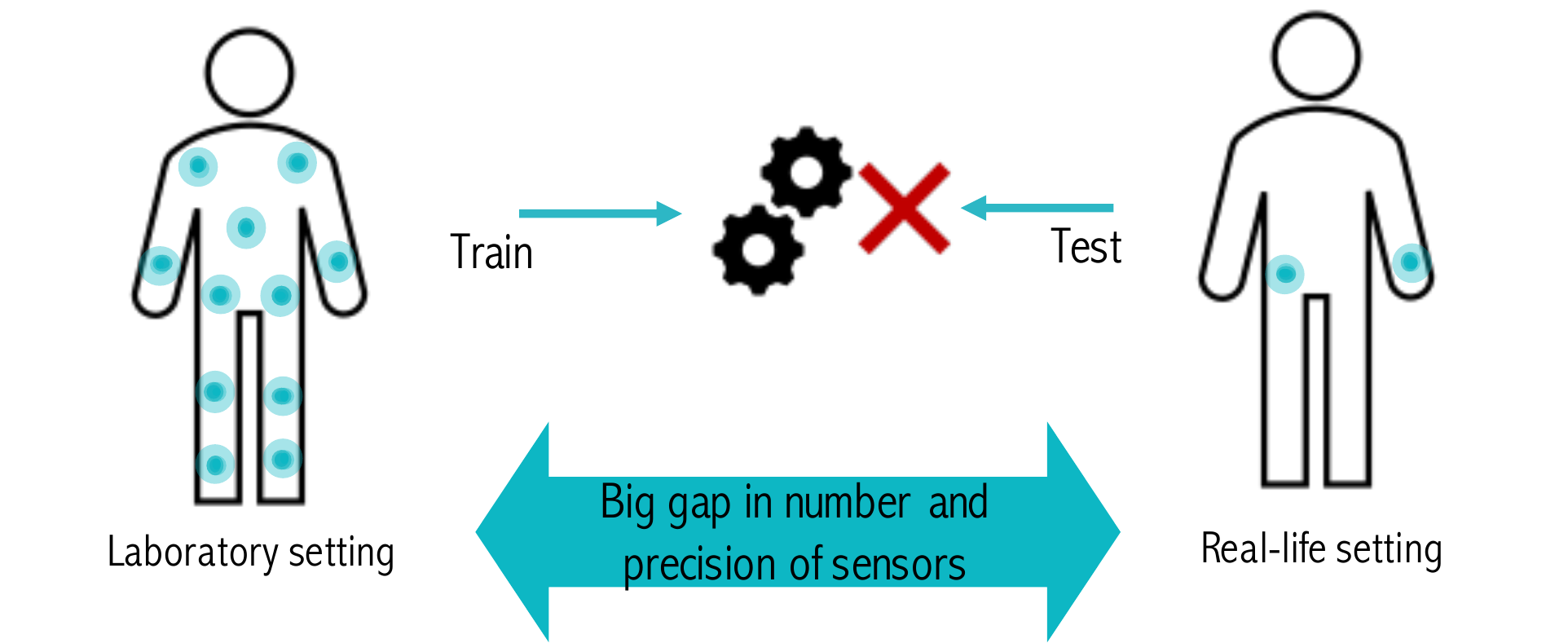}
    \caption{Multi-sensor activity recognition models are trained with data from multiple sensors. However, not all sensors may be available for the final user. If sensors are different during test time, the different input dimensionality will make the model unusable. }
    \label{fig:conceptual_figure}
\end{figure*}

Current plans to improve activity recognition accuracy usually rely on placing more devices in different body positions or using more sensors on the same device~\cite{Jain2018}.
Compared to using a single device, using sensors in multiple body locations can increase the accuracy by 35\% and by as much as 52\%~\cite{Bao2004}. Using multiple modalities on a single device can increase the accuracy by 25\% compared to using only the accelerometer sensor of a wearable device~\cite{tanzeem_practical}\footnote{These improvements are seen in physical activities and user-dependent models}. However, even if we can collect experimental data with highly accurate and multiple sensors, the results of these experiments are usually unrealistically high compared to those obtained in real life. This is because a typical user will not have all the same sensors used in the laboratory. As a consequence of this, \textit{models trained with experimental data are hard to transfer to real-life situations}. The difference in the number and precision of sensors used in training and real-life settings makes the models trained with laboratory data unusable in real-life~(Figure~\ref{fig:conceptual_figure}). Nevertheless, training models with only one sensor which is normally the case in real settings considerably reduces the performance of activity recognition as it does not take full advantage of the data collected for training.




In this paper, we study the possibility of building single-sensor activity recognition with multiple sensor training data. We formulate the learning problem and propose a method for its solution~(Section~\ref{sec:method}). Our approach uses representation learning and feature mapping to bring both the training and testing data to a shared representation space which is used as an input for the learning task. This solution is inspired by various transfer learning approaches~\cite{Raina2007,xu16dsne,Pan2008,Argyriou2006,Saeed2019} which learn a shared representation space from the source-task data and then apply it to the target learning task. Previous approaches, however, do not assume the setting of having different number of sensors between training and testing phases; rather, they take advantage of having more samples (unlabeled data, data with different labels or data from the same number of sensors but in different layouts). Our approach takes advantage of having additional sensors for learning. We further discuss the differences between our proposal and related work in in Section~\ref{sec:related}.
The main contributions of this paper can be summarized as follows:
\begin{enumerate}
    \item We formulate a problem in which multi-sensor data is available during the training phase of the activity recognition models, yet only a single sensor is available for testing. This problem setting represents the current gap between laboratory and real-life settings in complex activity recognition. We propose a method to solve this problem using representation learning and feature mapping, by which we can take advantage of multi-sensor data for training~(Section~\ref{sec:method}). The new space represents actions that are easier to recognize that the initial complex activities.
    \item We extensively evaluate the proposed method using four publicly available datasets with both simple and complex activities to assess the hypothesis that the method is more effective when dealing with complex activities. Our results show that the proposed method improves the recognition performance (F1-Score) for complex activities by as much as 17\%~(Section~\ref{sec:evaluation}).
    \item We use a boosting approach to further improve the performance which results in an improvement of 15\% compared to that of a method without  boosting~(Section~\ref{sec:evaluation}).
    \item We discuss why the proposed method improves performance for complex activities by analyzing the decomposition into actions. We also discuss further implications of this study including how to take advantage of settings with additional sensors for training~(Section~\ref{sec:discussion}).
\end{enumerate}

Our results show that we can achieve improvements by as much as 17 percentage points in the F1-Score in user-independent models~(Section~\ref{sec:evaluation}). The main advantage of the proposed approach is that it is suitable for any set of sensors and can work with any number of algorithms.
To the best of our knowledge, this work is the first to try single sensor-based activity recognition while training with multiple (high dimensional) sensors. Our solution combines three solutions studied before for different problems (feature learning, feature mapping and supervised learning) to leverage the strengths of the multiple-sensor data to improve single-sensor-based activity recognition.  We discuss the implications of our results, practical application scenarios and limitations of this study in Section~\ref{sec:discussion}. Finally, we present our conclusions in Section~\ref{sec:conclusions}.


\section{Related Work}
\label{sec:related}
Our work builds upon current research efforts to reduce the need of labeled data for activity recognition such as semi-supervised learning~\cite{Guan2007,semisupervised_tanzeem:2007,5254653,Bhattacharya2014,7398506,WEN201719,Stikic2011}, and transfer learning~\cite{VanKasteren2010, Blanke2010, Rashidi2011,Hu2011,adARC-Roggen2013,Cook2013}. These approaches take advantage of additional data; semi-supervised learning takes advantage of unlabeled data and transfer learning usually takes advantage of labeled data from a different domain by referring to an input of the same dimension but with a different probability distribution or data from a different task (data with the same dimension but with different labels)~\cite{transferlearningsurvey-pan2010}.

Our method is inspired in techniques that learn a shared, low-dimensional  representation for transfer learning~\cite{Argyriou2006,Raina2007,Pan2008, Saeed2019}. Like these works, we propose an approach in which we first learn a common representation for the learning task. Learned representations have been successful in activity recognition because they can be more robust to small variations in the input features rather than the hand-crafted features~\cite{Bhattacharya2014,Shirahama2016,Nguyen:2015:IDS:2750858.2804256,JingenLiu:2011:RHA:2191740.2191802, Shirahama2016,Plotz2011}. The underlying difference in our work is that we assume that the training data comes from a larger number of sensors compared to that of the testing data and that those sensors might be of different quality. We intend to take advantage of the experimental data where more sensors than those used in real life are available. Research in transfer learning has shown that the transfer is more successful when the tasks are related~\cite{transferlearningsurvey-pan2010}. In the problem setting, we are proposing that the relation comes from the fact that the training and testing input represents the same activities, thus the actions that compose them are the same.

In activity recognition, transfer learning has studied problems of source and target domains having different sets of activities, different sensor layouts or different sets of sensors~\cite{Cook2013}. Our proposal assumes that the source (training data) and target (testing data) domains have a different set of sensors. In~\cite{adARC-Roggen2013}, a teacher-learner approach was used to label the instances of the new set of sensors. However, the only knowledge transferred was that of the label to be assigned to the instance. Transfer learning based on feature representation have used manually designed features~\cite{VanKasteren2010} and learned features~\cite{Samarah2018, Xu2016}. Our approach is similar to those using a unified learned feature space (also called "latent variables"). However, these approaches assume the same number of features for training and testing because both training and test inputs are images~\cite{Xu2016}, or use the same set of sensors but in different layouts~\cite{Samarah2018}. This makes the mapping from the feature space to the shared representation space straightforward. Nevertheless, when a different set of sensors is used, its consequent mapping must also be learned. A similar work uses multi-task learning~\cite{Saeed2019} for learning a shared representation space across datasets. In this work, all datasets were collected with a single device, hence not considering the use of multiple sensors at different positions, or different dimensions of input when testing.

A setting with a similar solution approach is zero-shot learning where a semantic representation is used to describe activities, and the sensor features must be mapped to this representation~\cite{Cheng2013, Nguyen:2015:RNA:2802083.2808388}. In this setting, the mapping is learned by formulating it as a multi-label classification, or a multiple regression problem depending on whether the attributes are binary or continuous. Inspired by the usage of attribute mapping in zero-shot learning techniques~\cite{5206594}, we also use feature mapping to change data representation from the single sensor feature space to a learned space that represents information from multiple sensors. However, the problem setting of zero-shot learning is different from the problem setting studied in this paper.

Our proposed method takes advantage of data from additional sensing sources that complement the training data. In this regard, our approach can be thought as an instance of domain adaptation~\cite{xu16dsne, Natarajan2016}, where the training data is different from the testing data. However, the case where this difference becomes additional information sources as studied in this work, has attracted less attention. Although this problem setting of using higher dimensional data for training is similar to the Learning Under Privileged Information paradigm~\cite{Vapnik2015}, this paradigm uses privileged information to accelerate the learning rate for specific algorithms whereas we propose a generic approach that can be used with any classification algorithm. Using this approach, the representation space can be more robust and the labeling effort can be reduced.

In summary, the main difference between our problem setting and other problem settings is that we assume that training data differs from the test data; not only in the number of sensors (features), but also in the quality of the sensors used.  This difference forces us to develop a new method of learning as current classification models assume that the dimensions of the input does not change in testing.  We propose an approach based on feature learning and feature mapping from single sensor features to learned features from multiple sensors. The differences between our proposed approach and other approaches are:
\begin{enumerate}
    \item we do not assume any knowledge about the attributes that describe each activity, similar to attribute-based learning approaches and feature transfer learning approaches. We use high-dimensional data to learn this.
    \item Given the different input dimensions between training and testing data, mapping test data to learned features is not as straightforward as it is in other feature-learning approaches. We use a multiple-regression approach for this mapping.
    \item The proposed approach can use any set of sensors and can adapt to suit any combination of algorithms.
\end{enumerate}
\section{Single-Sensor Activity Recognition from Multi-Sensor Training Data}
\label{sec:method}
\begin{figure*}
    \centering
    \includegraphics[width=0.65\textwidth]{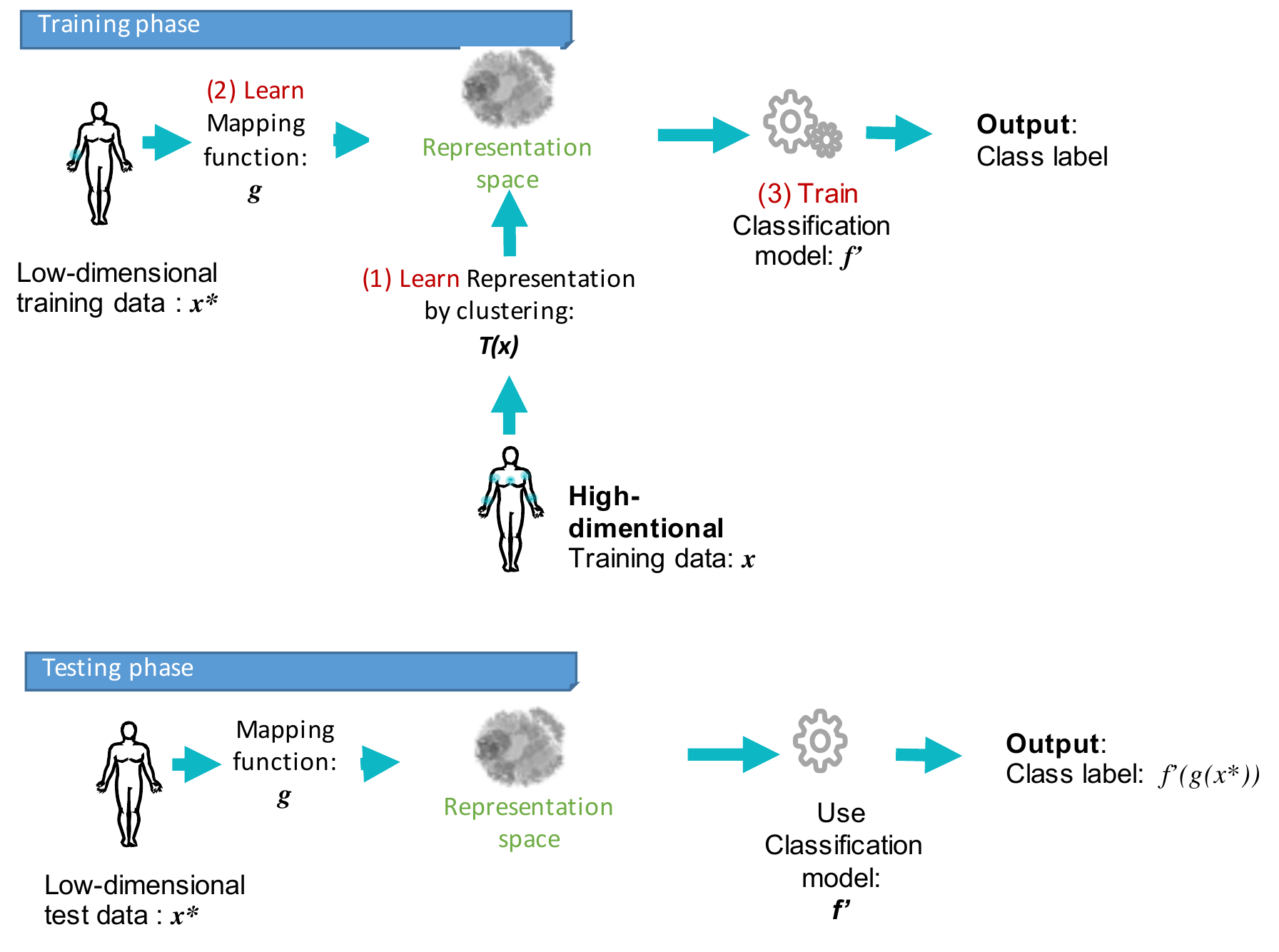}
    \caption{Proposed algorithm to learn a single-sensor activity recognition model using multi-sensor data. During the training phase (top) multiple sensors are used to learn a representation space (1). A mapping from single sensor to the representation step is learned (2), and the classification model is trained with the representation space as inputs (3). The final model $f'(g(x^*)$ is used for testing. }
    \label{fig:solution_figure}
\end{figure*}

In this section, we describe our proposed approach.
This method uses knowledge that can be obtained from data in controlled settings using multiple sensors in settings where fewer and possibly less precise sensors are being used. Such condition causes inputs of different dimensionality for training and testing.
We first present an overview of the method and challenges (Section~\ref{sec:overview}), then we define the problem formally (Section~\ref{sec:problem}) and the proposed algorithm  (Section~\ref{sec:algorithm}).

\subsection{Overview}
\label{sec:overview}
To learn using more sensors than those available in the final system, we have to overcome the main challenge which is to find a common representation for the training and testing data. This is because classification algorithms assume no change for the dimensions of the input during testing. Therefore, we use representation learning and feature mapping to solve this challenge~(Figure~\ref{fig:solution_figure}). Representation learning is used to learn features that encode information from the whole-body sensors (or multiple sensors). The information provided by multiple sensors help us to discern activities that may have similar movements in one limb but have different movements in other limbs. Feature mapping is used to map the features from the single-sensor to this learned representation.

\begin{figure}
    \centering
    \includegraphics[width=0.45\columnwidth]{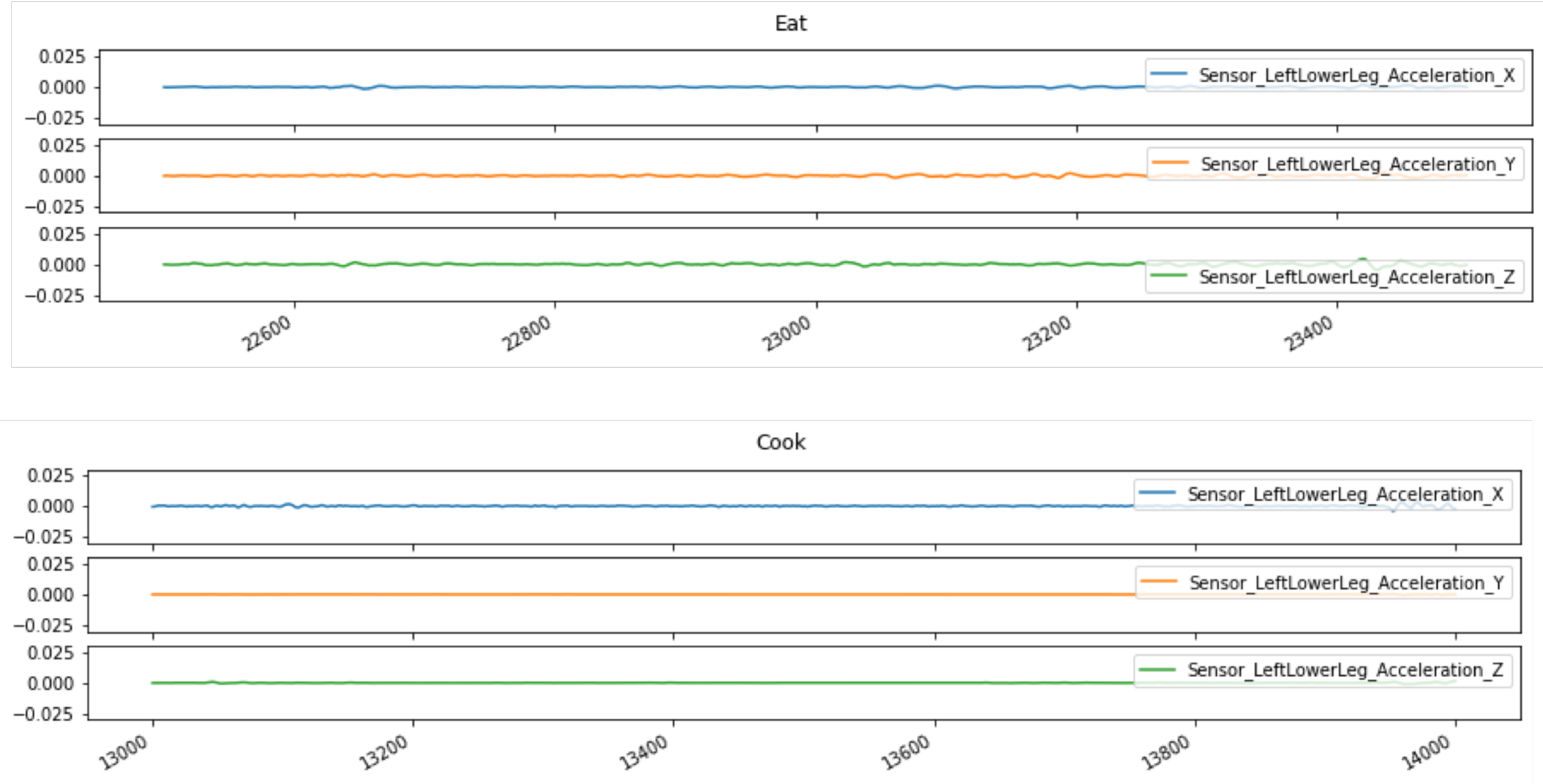}
    \includegraphics[width=0.45\columnwidth]{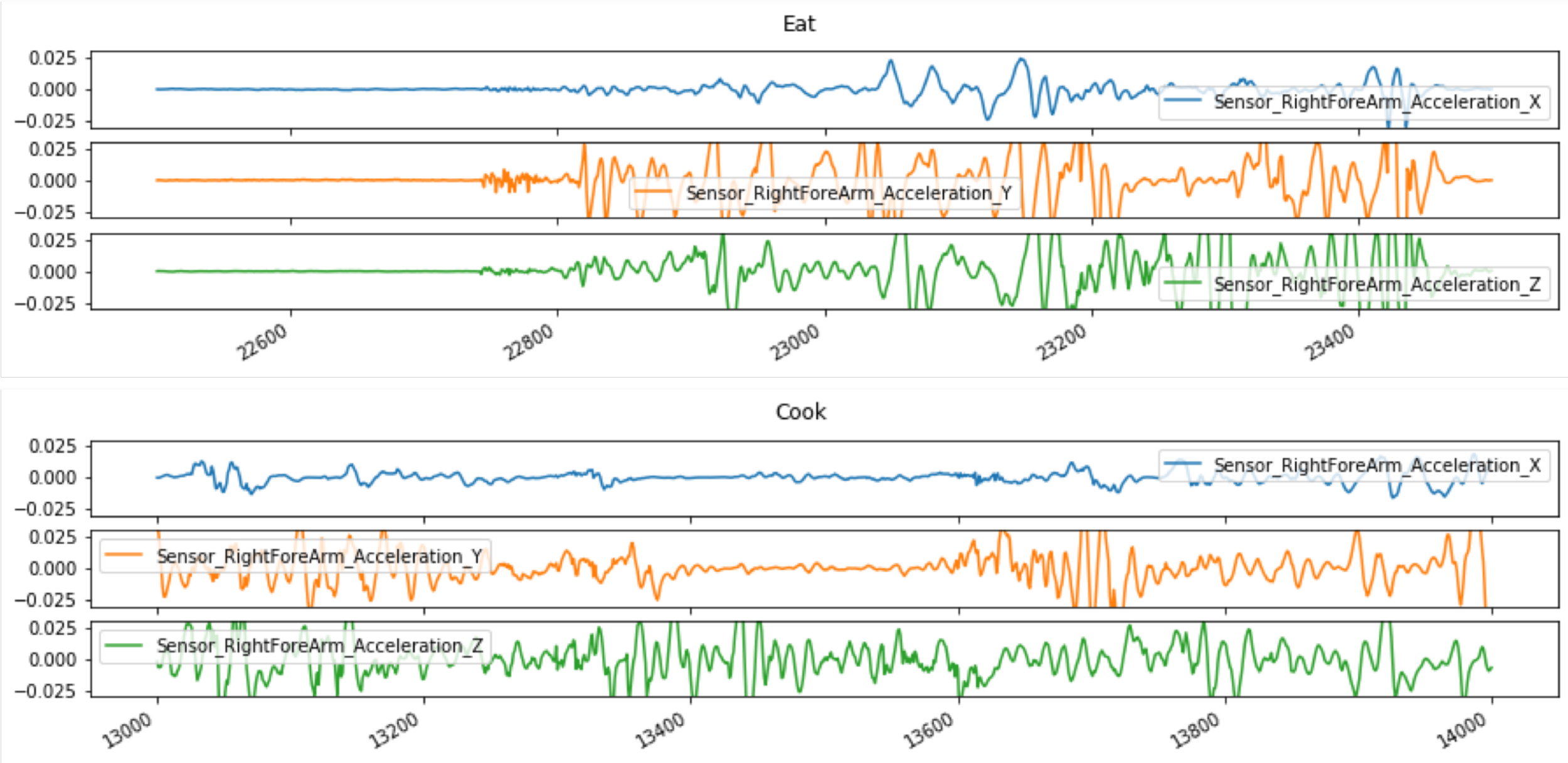}
    \caption{Differentiating "eat" from "cook" is difficult with only the leg sensor (top) but the arm sensor (bottom) can provide hints of the differences}
    \label{fig:eat-cook}
\end{figure}

\begin{figure}
    \centering
    \includegraphics[width=0.45\columnwidth]{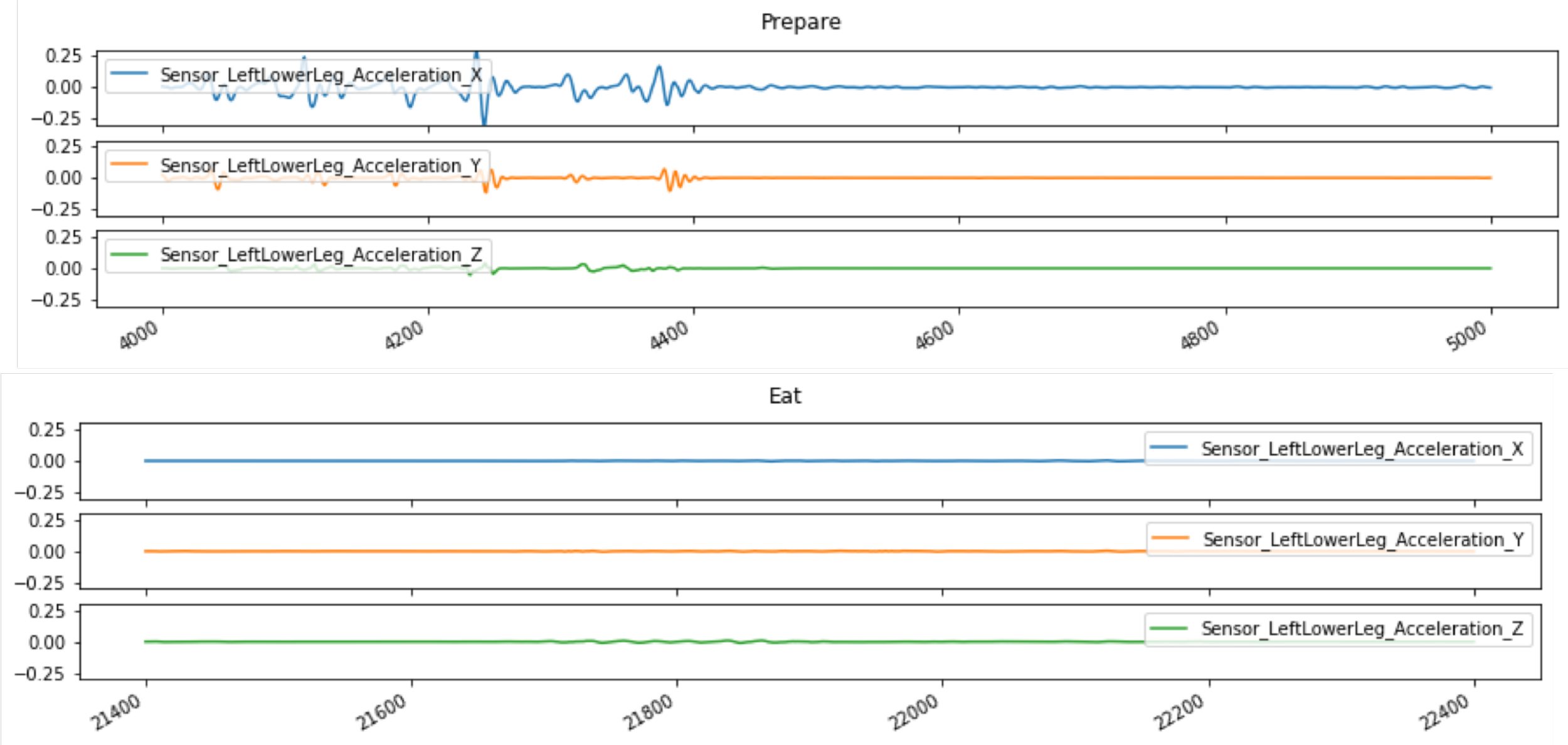}
    \includegraphics[width=0.45\columnwidth]{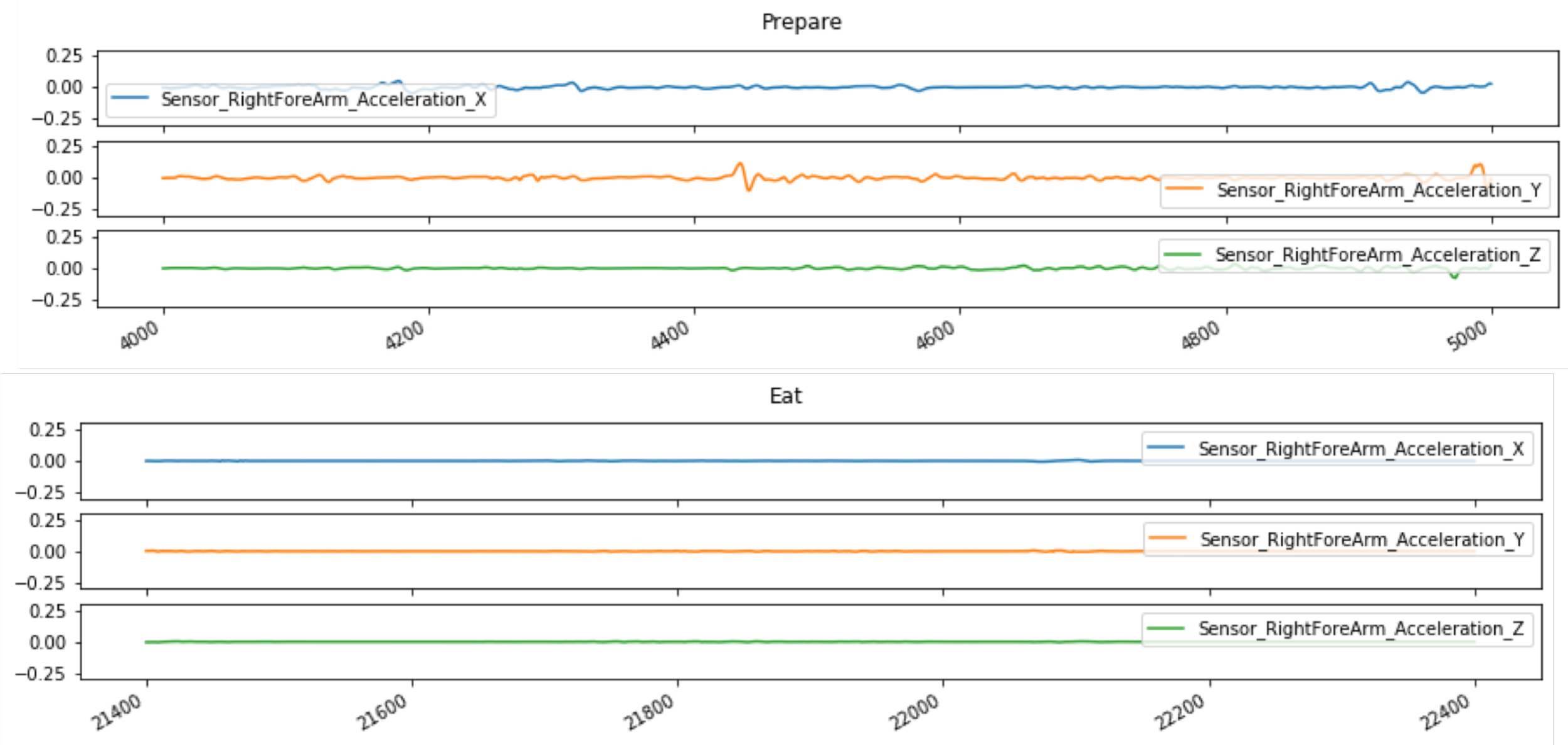}
    \caption{In contrast, to differentiate "prepare" from "eat", the leg sensor (top) might provide more information than the arm sensor (bottom)}
    \label{fig:prepare-eat}
\end{figure}

To understand why combining multiple sensor information into a single representation is helpful, let us consider the examples in Figures~\ref{fig:eat-cook} and \ref{fig:prepare-eat} where windows of different activities are depicted. As noticed in the figures, for the pair eat-cook, the sensor in the arm discerns better between both activities whereas, for the prepare-eat pair, the leg sensor might discern better. However, using the sensor in the wrist is more practical so when the user wears the sensor in the arm position for all activities there might be confusions. The learned representation, combining the information from both arm and leg, can help to discern those activities that were initially thought to be the same. The second challenge is to correctly map the single sensor to the learned representation. This is achievable because we map independently to each of the features in the new space, which are simpler classes compared to the complex activities. The new features can be thought of as actions.

After learning the representation space, the next challenge is to learn to classify activities (Step 3 in Figure~\ref{fig:solution_figure}). This is a traditional supervised learning problem but given the problem setting, data for this model can come from multiple sensors or from the single sensor. In the testing or usage phase, the mapping function and classification model are used to recognize activities.

\subsection{Problem Formulation}
\label{sec:problem}
The proposed problem setting is one where the training and testing data have different dimensionality.
Training data has higher dimensionality than that of testing data because it is measured from \emph{multiple sensors} whereas testing data is measured with a single sensor. Moreover, we can assume that this sensor may have lower accuracy and precision than the sensors used in the training data.

We assume that a feature space exists that represents the non-redundant and important information needed to classify the activities; the \emph{representation space}. The first step in the algorithm is to learn this feature space from the multi-sensor training data. In this paper, we use clusters of simple sensor features as the representation space; however, other representation learning algorithms can be used. Our main goal in this paper is to evaluate the feasibility of this approach rather than evaluating feature learning approaches.

Let us now describe the problem setting as compared to traditional activity recognition.
The activity recognition problem is typically defined as a classification problem.
We train an activity recognition model using a set of pairs ($x, y$), where $x$ is a feature vector and $y$ is an activity label. The task of the traditional activity recognition is to find a function $f(x)$ that minimizes the classification error $f(x) \neq y$.

In the proposed problem setting, we have a set of triples ($ x^*, x, y$), where $x^*\in R^m$, $x \in R^n$,  and $m \ll n$.
The task is to find functions $f'$ and $g'$ so that the classification error $f'(g(x^*)) \neq y$ is minimized. The \emph{representation space} is learned from $x$ as $T(x)$

\subsection{Learning Algorithm}
\label{sec:algorithm}
Using the previous problem formulation, we now describe the proposed algorithm~(Algorithm~\ref{alg:solution}).

We use unsupervised learning for feature learning as it enables the use of a high volume of unlabeled data. Therefore, the input of the algorithms are three sets of data instead of a triple as stated in the problem formulation; one consisting of an unlabeled multi-sensor data, one pair of multi-sensor and single sensor data, and a pair of labeled single sensor data.

The algorithm first learns the representation space (Line~\ref{line:rep}) using multi-sensor data. Representation-learning algorithms include dimensionality reduction algorithms such as PCA, and clustering and supervised learning algorithms like encoders. In this paper, we use feature clustering.
The goal is to find a representation space of dimension $d$, with $n>d>m$, so that it acts as a middle layer between the multi and single sensor data spaces.

\begin{algorithm}[H]
\LinesNumbered
\KwData{($x_1)$ ($x_2$, $x^*_1$), ($x^*_2$, $y$), $x\in R^n, x^*\in R^m, d$}
\KwResult{Mapping function and activity classification model}
 \tcc{Learn representation}
 repModel = learnRepresentation($x_1$, $d$))\\ \nllabel{line:rep}

 mappings = [] \\
 repData = repModel($x_2$)\\
 i = 0\
 \While{$i < d$}{ \nllabel{line:mapping_s}
       target = repData[:, column]\;
       mapping[column] = linReg($x^*_1$, target) \nllabel{line:mapping_e}
 }
 \tcc{Classification model}
 data = []\\
 i=0\\
 \While{$i < d$}{
       data = data.append(mapping[column]($x^*_2$))\
 }
 model = learn\_classification(data, $y$)\\
 \label{alg:solution}
 \caption{Learning algorithm using multi-sensor data for training and single-sensor data for testing}
\end{algorithm}

The next step is to learn a mapping that takes the single-sensor data to the middle-layer representation space. We use multiple regression for this (Line~\ref{line:mapping_s} to Line~\ref{line:mapping_e}). In this paper, we evaluate both linear and logistic regression for this step.

Notice that the dataset used for learning the representation ($x_1$) can differ from that used to learn the mapping ($x_2$).
Moreover, notice that both datasets are unlabeled as only data from the sensors are needed for these two first steps.

Finally, the classifier is learned using features of the learned representation space as input and activity labels as output. This is equivalent to a traditional supervised machine learning problem for activity recognition. Notice that this is the only step where labeled data is needed.

The main advantage of having separate processes for learning the feature space, mapping and classifier is that we can use a small dataset to learn the final classifier while using a large unlabeled dataset for learning the representation and mapping.
Since it is common to collect more unlabeled data than labeled data (free-style experiments can be conducted for obtaining unlabeled datasets), we can use data that covers a wide range of contexts for steps 1 and 2, making the representation space more robust.

This algorithm has been implemented in Python for the evaluation.

\subsection{Leveraging Models for Performance}
The second challenge in the proposed approach aims at minimizing the final classification error.
Given the serial architecture of the proposed method, the final classification error will depend on the mapping and classification errors. In short, even a perfect classifier can have a high error for certain activities if the mapping to the common representation space is not perfect.

We propose to combine the traditional activity recognition model with the proposed model to achieve a better performance.
The basic intuition is that the proposed model will help to better differentiate difficult activities from each other whereas other activities will be well recognized by the traditional approach.
We use a multi-class adaboost algorithm~\cite{hastie2009multi} to combine the two learners; proposed method and traditional approach.
For this, we first train the activity recognition model using the proposed method. Then, based on the training errors of this model, we train a classifier using the traditional approach giving more weight to the samples with high error that are identified in the first step.

\subsection{Effective Domains of the Proposed Method}
\label{sec:expected}
We hypothesize that, for our method to work,  each feature $f_i$ in the learned representation space $T(x)$ should be equivalent to an action or a low-level activity. Recognition of low-level activities is then easier from a single sensor.
Given that physical activities can be easily recognized from a low-precision sensor, their feature learning will not improve the total accuracy or equivalent, compared with that of the traditional method.
On the other hand, for complex activities, we can expect that the mapping from single-sensor features $x$ to the learned representation space $T(x)$ will be effective, and the activity classification from the learned representation space $T(x)$ to the complex activities $y$ will be better than that from single sensor features.

Given this rationale, we expect that the proposed method is mainly effective when dealing with complex activities, which can be decomposed into actions.
Besides, for physical and low-level activities, the proposed method can keep equivalent or not worse performance to traditional methods. We will demonstrate these assumptions in the following section.
\section{Experimental Evaluation}
\label{sec:evaluation}
In this section, we describe the experimental settings and results of the evaluation for the proposed approach. The evaluation compares the performance of the proposed method against a traditional pipeline using a single (and the same) sensor for both training and testing. Since our method uses clustering for learning the representation space, we first evaluate the performance of a classifier using these clusters as features. This becomes the highest expected performance of our method.

We expect the method to be more useful for complex activities as these activities have shown higher improvements when using multiple sensors. We also expect it to have a higher impact when the training sensors have higher quality than the testing sensors as they can learn a stronger representation.

Furthermore, we evaluate the sensitivity of the proposed approach to two parameters, namely the window size parameter and the number and type of sensors included for training.

In the following section, we describe the datasets used for the evaluation~(Section~\ref{sec:dataset}), implementation and evaluation scenarios (Section~\ref{sec:eval_setup}) and summarize the results obtained in the performance evaluation~(Section~\ref{sec:results}).

\subsection{Datasets}
\label{sec:dataset}
To evaluate our method, we consider four publicly available datasets with several sensors in different placements. Some important aspects of the data are summarized in Table~\ref{tab:datasets}. We classify the activities as complex, gestures, and physical activities. Complex activities have longer durations and are composed of several different actions. Physical activities, considered simpler, involve the repetition of an action, possibly periodically. For example, \textsc{walk} involves the repetition of steps. Gestures are short and may or may not be repetitive; for example, \textsc{take} consists of a non-periodic movement but \textsc{Cut} may involve repetition of movement.  Below, we describe briefly each dataset, summarize its key points and the data pre-processing used.

\begin{table*}
    \centering
    \renewcommand{\arraystretch}{1}
     \small{
    \begin{tabular}{p{0.2\textwidth}p{0.1\textwidth}p{0.1\textwidth}p{0.1\textwidth}p{0.1\textwidth}p{0.2\textwidth}}
        \toprule\\
         Dataset (Activity type) &  \textnumero~of sensors &  \textnumero~of~subjects &  \textnumero~of~classes &  \textnumero~of~windows & Locations (sampling rate)  \\
         \midrule\\
         OPP HL~\cite{OPP_Dataset} (complex activities) & 5 IMUs & 4 & 6 & & Upper and lower arms and back (30Hz) \\
         Cooking dataset~\cite{cookingdataset} (complex activities) & 5 IMUs & 7 & 16& 7105 & Lower legs, Lower arms, and upper back (120 Hz)\\
         PAMAP~\cite{PAMAPDataset} (physical activities)& 3 IMUs & 9 & 12 & 28720 & Chest, dominant hand and ankle (100Hz)\\
         OPP Locomotion~\cite{OPP_Dataset} (physical activities)& 5 IMUs & 4 & 5 & 21013 & Upper and lower arms and back (30Hz) \\
         \bottomrule
    \end{tabular}}
    \caption{Datasets used for the evaluation}
    \label{tab:datasets}
\end{table*}

\subsubsection{Cooking Task Dataset:}~\cite{cookingdataset} This dataset recreates a meal-time routine consisting of the following major tasks: (i) Prepare a soup (ii) Set table. (iii) Eat meal. (iv) Clean up and put away utensils. The dataset is labeled with gestures, and we set the labels for these 4 activities by analyzing the activities done and the script given in the paper. The dataset was collected in a laboratory setting with some utensils replaced by physical props and some actions shortened. We use windows of 1 second with 0.25sec step. In this evaluation, we used the accelerometer sensor of the IMU sensors.

\subsubsection{Opportunity Dataset}
The Opportunity dataset~\cite{OPP_Dataset} was recorded in a room simulating a studio and includes activities of a morning routine performed by 4 subjects.
The activities were labeled in different levels; locomotion, gestures, and high-level activities. In our evaluation, we used the high-level activities (OPP HL) which include \textsc{Relaxing}, \textsc{Coffee}  (prepare and drink), \textsc{Sandwich} (prepare and eat), \textsc{Early-morning} (check objects in the room), and \textsc{cleanup}, and the locomotion activities (OPP Loc) which are: \textsc{Stand},\textsc{Walk},\textsc{Sit},\textsc{Lie}, and \textsc{Unlabeled}.
In total, there are almost 6 h of recorded activity in this dataset plus an additional 2 h of unlabeled data ('Drill' run). We use this unlabeled data in the feature learning stage.

Although the dataset includes multiple on-body and object sensors, in this evaluation we used only the accelerometer sensor of the IMU sensors. These sensors are only placed in the upper body thus, the placements are not as diverse as in the cooking dataset.

For the high-level activities, we used windows of 30 s with 15 s step.  For the locomotion activities, we used 3-s windows with a 2-s step (1-s overlap).

\subsubsection{PAMAP Dataset}
The Physical Activity Monitoring Dataset~\cite{PAMAPDataset} (PAMAP) is a benchmark dataset for physical activity monitoring.
The activities are; \textsc{lie, sit, stand, walk, run, cycle, Nordic walk, iron, vacuum clean, rope jump, and ascend and descend stairs}. Although some subjects performed other activities during data collection, we did not include them as not all subjects performed those activities during that time. This dataset contains approximately 8 h of recorded activities. We used windows of 5.12 s with a step of 1 s as per recommendations of the original publication.

\subsection{Implementation and Experimental Setup}
\label{sec:eval_setup}
We implemented the algorithm proposed in Section~\ref{sec:method} in Python using the scikit-learn library for the evaluation.  We used feature clustering to learn the feature space using 3 clusters per sensor in all cases. That is, we use 15 clusters for datasets using 5 sensors, and 9 clusters for PAMAP.
We implemented the algorithm with and without boosting, using linear and logistic regression for the mapping function.
In total, we implemented and evaluated 4 algorithms that depended on the mapping function and if the boosting approach was used or not: linear regression (LinR), logistic regression (LogR), linear with boosting (LinB) and logistic with boosting (LogB).

For each sensor axis, we extracted the following statistical features; mean, standard deviation, range (the difference between minimum and maximum) and the difference between the mean and the median. We used the same features for all datasets.

As \textit{ evaluation protocol}, we used a leave-one-subject-out cross-validation approach (user-independent models). This approach evaluates the robustness of the classifier to new users. We chose this approach as our method is intended to transfer controlled experiments into real-life settings where users will be unseen. In this protocol, we used data from all users except one for both training stages, which is both feature learning and mapping learning.

As \textit{evaluation metrics}, we used the micro average F1-Score~\cite{yang1999evaluation} to compare the performance across all activities of the proposed and traditional approach.

\subsection{Results}
\label{sec:results}
We now present the results of the empirical evaluation. We first evaluate the hypothesis that states: using learned features from multiple sensors achieves better performance than traditional single-sensor features~(Section~\ref{sec:potential}). We then evaluate the effectiveness of the proposed method by assessing the performance when the single sensor is used as the input for feature mapping~(Sections~\ref{sec:effective_complex}, \ref{sec:effective_simple}, and \ref{sec:effective_boost}). Our results show that the improvement is proportional to the difference between the multi-sensor performance and the traditional single-sensor approach. Furthermore, we go on to evaluate how the proposed method performs when using lower-quality sensors for testing~(Section~\ref{sec:lower_quality}) and when including different types of sensors for learning the representation~(Section~\ref{sec:more_sensors}). The results are positive and encouraging, showing that the proposed method can take advantage of these settings.

\subsubsection{Measuring The Gap between Single-Sensor and Multi-Sensor Performance: Analysis of maximum expected improvement }
\label{sec:potential}

\begin{figure}
    \centering
    \includegraphics[width=0.9\columnwidth]{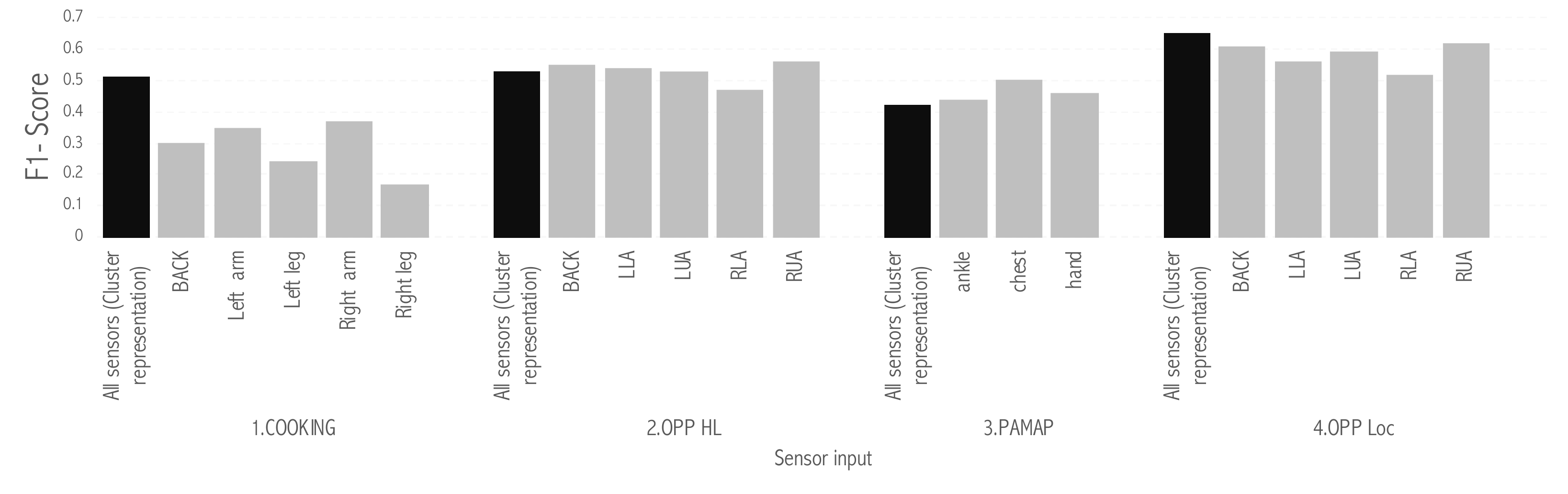}
    \caption{Preliminary analysis of the maximum expected performance compared to traditional approach with single sensor. The performance obtained when multiple sensors are used for training and testing (cluster representation) is the upper limit for the performance expected with the proposed approach.}
    \label{fig:potential}
\end{figure}

We first evaluate the gap between the performance of activity recognition using a single sensor and the performance of when learned features from all sensors are being used (cluster features). The main objective in this evaluation is to measure the maximum possible improvement, as the expected performance of the proposed approach will be less than the performance when all sensors are used also in testing.
As shown in Figure~\ref{fig:potential}, there is a potential for improving 34 percentage points the F1-Score of activity recognition using the right leg sensor in the Cooking Dataset (0.51 - 0.17), and a potential of only 0.09 when using the Left Lower Arm (LLA) sensor in the OPP Loc dataset (0.65 - 0.56). We can also observe cases where the performance of using clustered features is lower than that of using the individual sensors (PAMAP dataset).

The previous evaluation confirms that in most cases, and specially for complex activities, a classifier using multiple sensors (with learned features) in both training and testing achieves a better performance than one using a single sensor in both phases. We now evaluate the performance of the proposed approach, which uses multiple sensors for training and a single sensor for testing. The performance depends on an effective map learning and leveraging of both models when the boosting approach is used.

\subsubsection{Comparing the Proposed Approach against Traditional Single-Sensor on Complex Activities}
\label{sec:effective_complex}
As mentioned in Section~\ref{sec:expected}, we expect the proposed approach will yield the largest improvements for complex activities because they can be decomposed into actions when the features are learned.

\begin{figure}
    \centering
    \includegraphics[width=0.8\columnwidth]{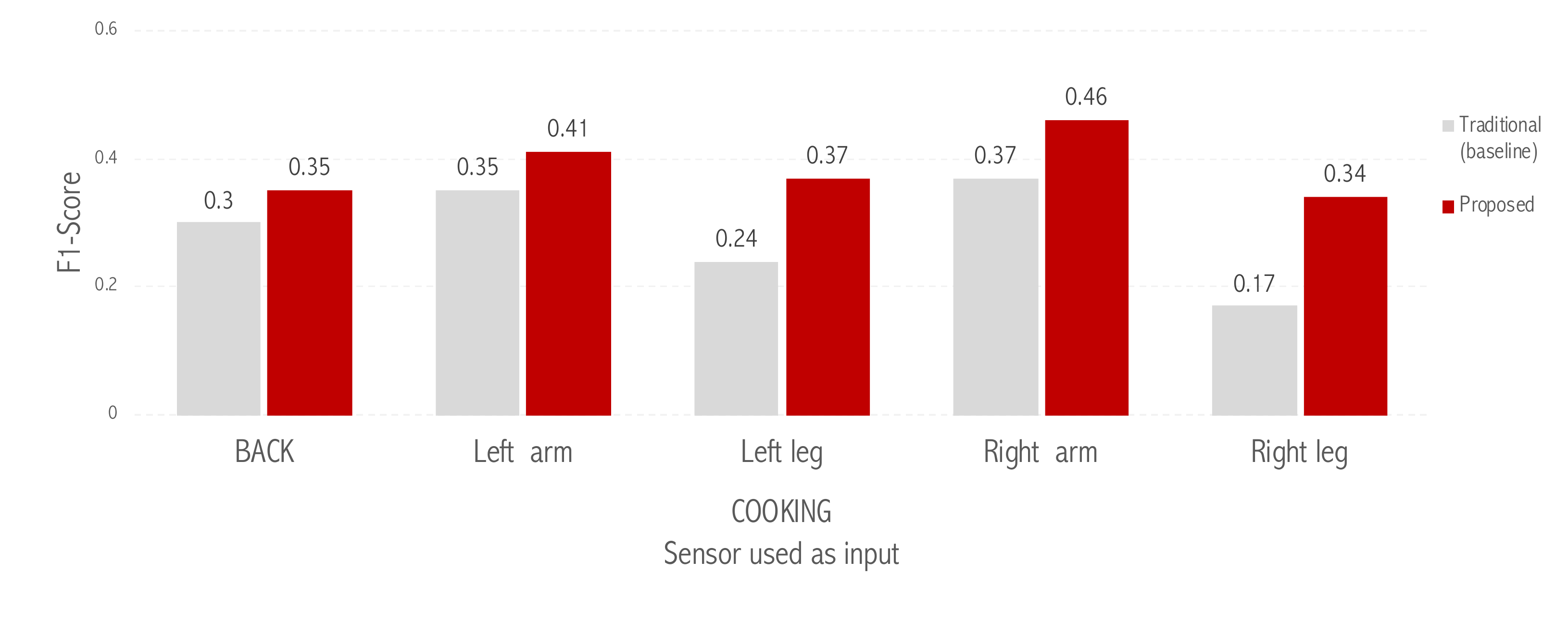}
    \caption{F1-Score of the proposed approach compared to the traditional approach using only one sensor as input in the Cooking Dataset. Improvements of 17\% and 13\% are observed for the right and left leg respectively. For the arms, improvements of 9\% and 6\% are observed. The improvements are proportional to the large differences between the traditional single and multi sensor activity recognition.}
    \label{fig:cooking-dataset}
\end{figure}

Figure~\ref{fig:cooking-dataset} shows the performance of the proposed approach compared to the baseline of a traditional approach using a single sensor in the Cooking dataset. It should be noted that we used the same features for the baseline and for learning the mapping in the proposed approach. The improvements in the proposed approach come from the use of better representation, learned using multiple sensors for the representation learning task. \emph{The observed improvements in the F1-Score are \textbf{17} percentage points (pp) for the right leg sensor, \textbf{13} pp for the left leg sensor, \textbf{9} pp for the right arm sensor, \textbf{6} pp for the left arm sensor and \textbf{5} pp for the back sensor}. These are significant improvements that show that we can leverage the use of multiple sensors when collecting training data for the models, while still using a single sensor during the usage of an activity recognition application.

\begin{figure}
    \centering
    \includegraphics[width=0.8\columnwidth]{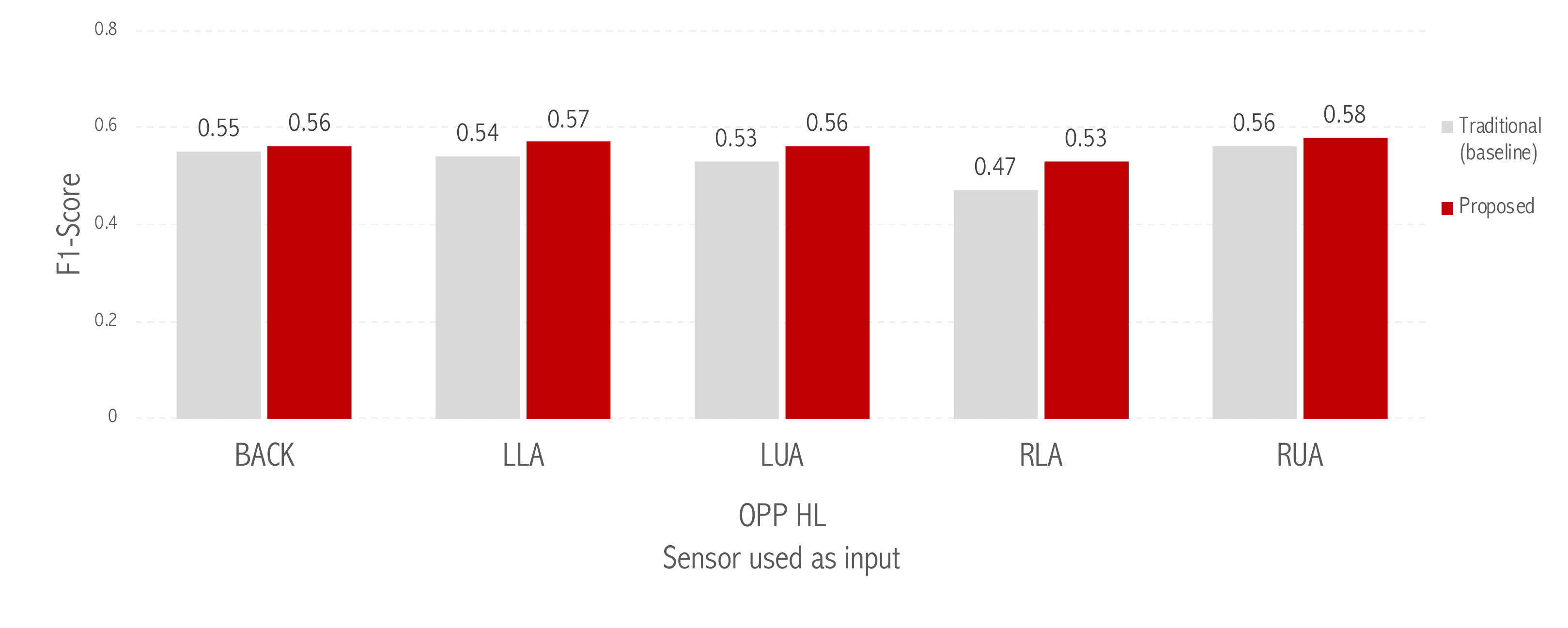}
    \caption{F1-Score of the proposed approach compared to the traditional approach using only one sensor as input in the OPP-HL Dataset. Improvements of 6\% and 3\% are observed for the RLA, LLA and LUA sensors. Again, these improvements are proportional to the differences between multiple and single sensor performance in Figure~\ref{fig:potential}.}
    \label{fig:opphl-dataset}
\end{figure}

In the Opportunity-HL dataset~(Figure~\ref{fig:opphl-dataset}), we observe the largest benefit from the proposed approach when there is a larger difference between the performance of the single-sensor compared to the multiple sensor in the traditional approach~(Figure~\ref{fig:potential}). The \emph{RLA sensor shows an improvement of 6 percentage points} (pp) in the F1-Score compared to the traditional single-sensor approach. Other sensors show improvements of 3pp (LLA and LUA), 2pp (RUA) and 1pp (BACK). This result is consistent with a smaller room for improvement observed in Figure~\ref{fig:potential}.

\subsubsection{Comparing the Proposed Approach against Traditional Single-Sensor on Physical Activities}
\label{sec:effective_simple}
\begin{figure}
    \centering
    \includegraphics[width=0.45\columnwidth]{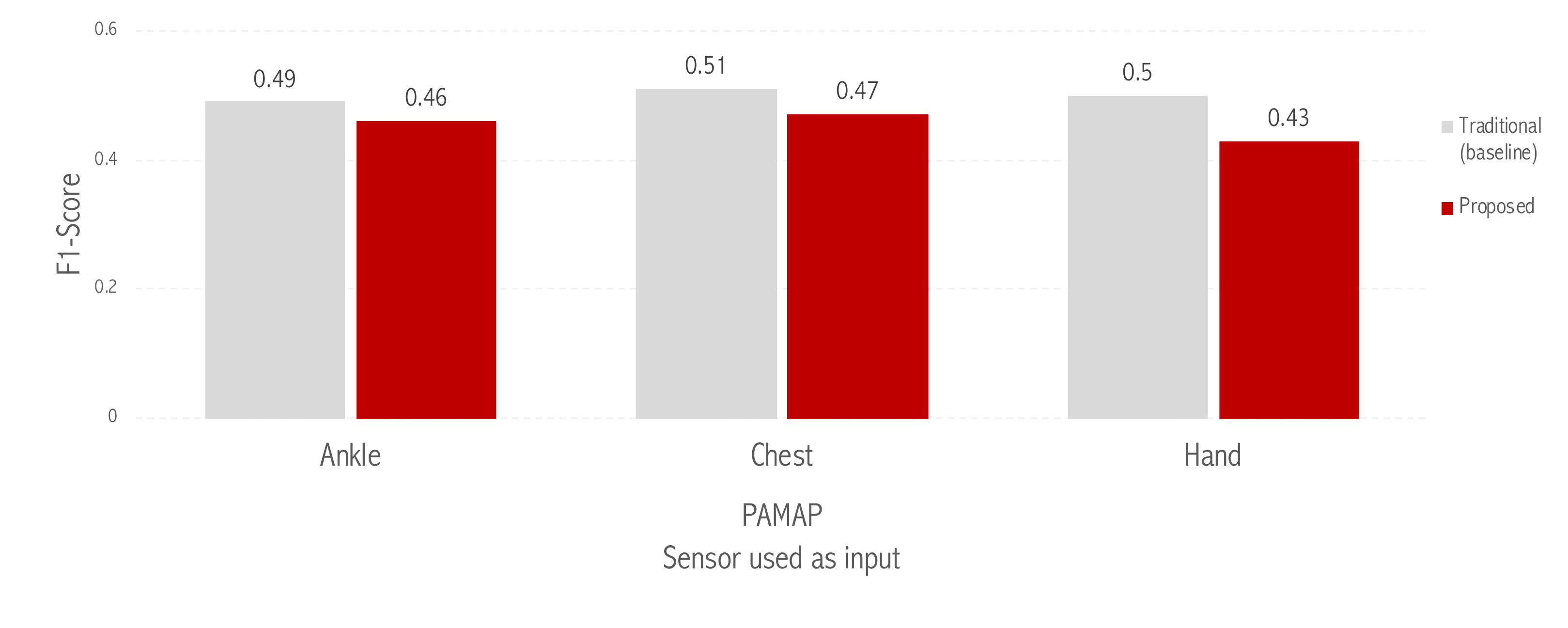}
    \includegraphics[width=0.45\columnwidth]{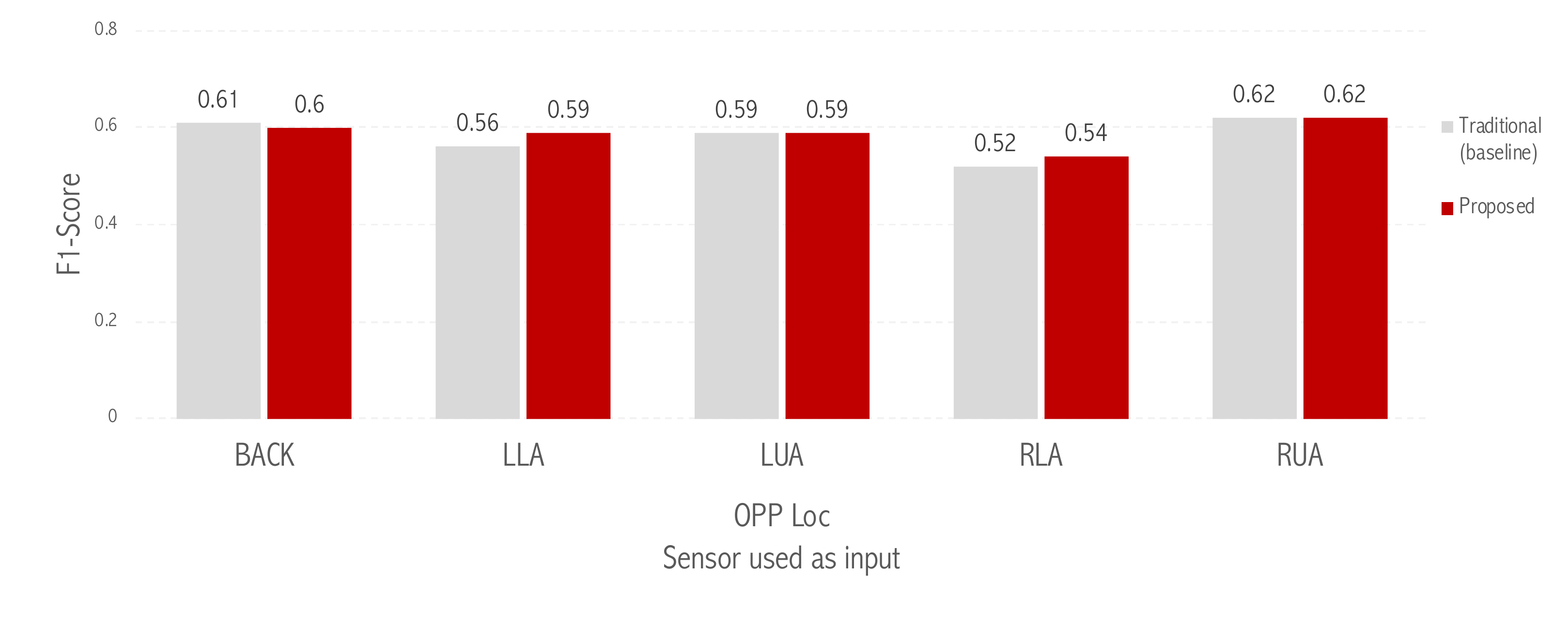}
    \caption{F1-Score of the proposed approach compared to the traditional single-sensor approach for simple activities (OPP Loc and PAMAP datasets). As was expected, for simple activities there are little to no improvements because they are not decomposed into smaller actions. }
    \label{fig:allddatasets}
\end{figure}

Figure~\ref{fig:allddatasets} shows the performance of the proposed approach compared to the baseline for datasets representing physical activities. In contrast to complex activities~(Section~\ref{sec:expected}), when we deal with physical activities~(OPP-Loc and PAMAP datasets), the performance of multiple sensors is similar or lower to that of the single sensor. Thus, the performance of the proposed approach is lower than that of the traditional approach.

\subsubsection{Performance of Boosting and no-Boosting Approaches}
\label{sec:effective_boost}
\begin{figure}
    \centering
    \includegraphics[width=0.95\columnwidth]{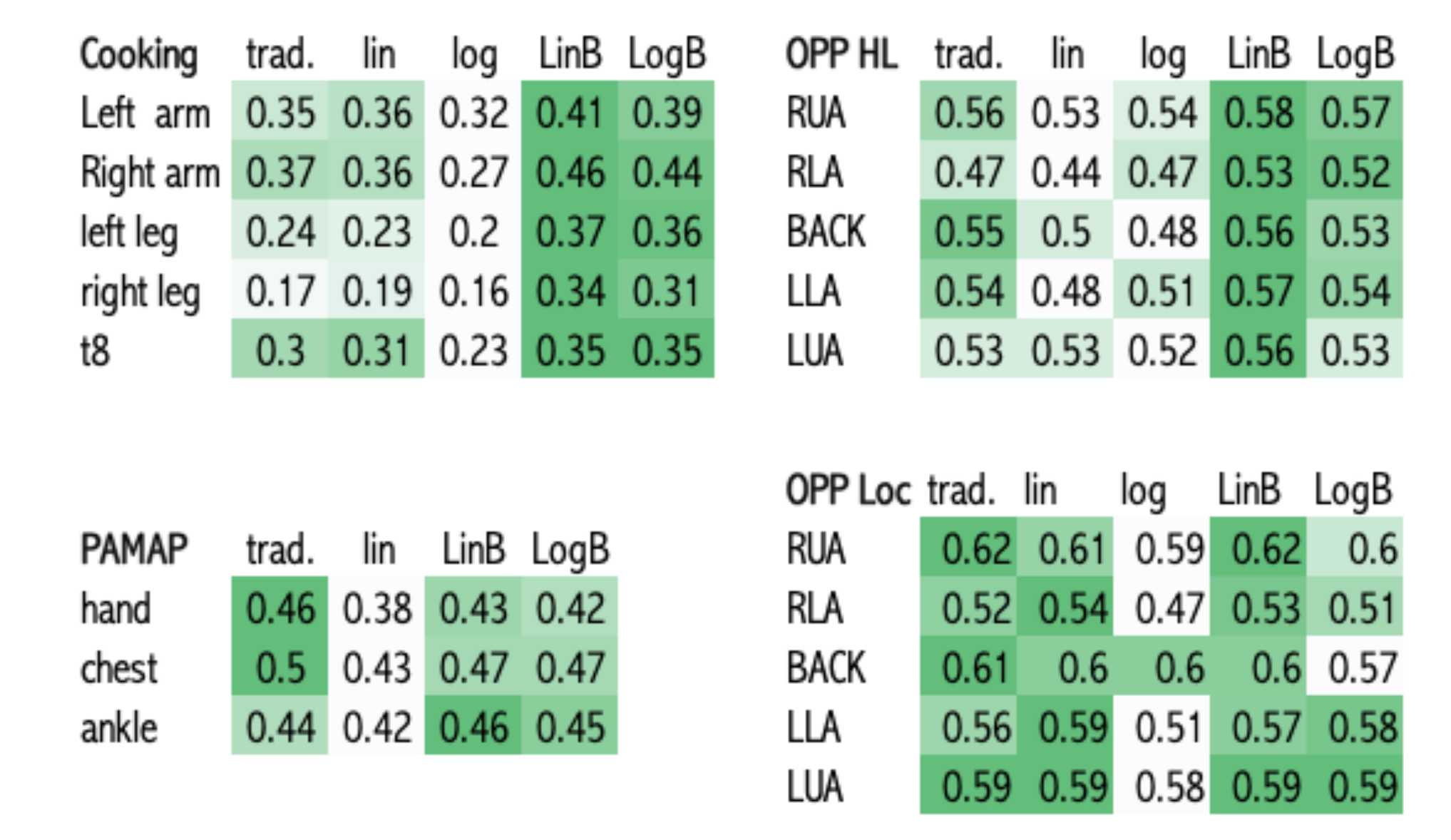}
    \caption{\textbf{F1-Score for the 4 algorithms implemented:} The best score is shown in Dark green background, showing that in complex and middle-level activities, the proposed methods always have the best performance. Linear mapping has better performance than logistic performance in general. In most cases, boosting further improves the performance.  }
    \label{fig:all_approaches_report}
\end{figure}

Figure~\ref{fig:all_approaches_report} shows the results for all the sensors using all four algorithms; linear regression (LinR), logistic regression (LogR), linear with boosting (LinB) and logistic with boosting (LogB). Even if the proposed method with linear mapping and boosting have the best performance in most cases, in the OPP-Loc dataset, the best performance is achieved when no boosting is used~(except for RUA sensor). In other cases, we can observe that using boosting improves the performance of the model by as much as 15 percentage points when compared to no boosting (refer to the right leg sensor in the cooking dataset).

\subsubsection{Performance when Testing Sensor has a Lower Quality than Training Sensors}
\label{sec:lower_quality}
The OPP dataset includes accelerometer sensors of two different quality levels. As specified by the dataset description, the accelerometer sensors can have more data loss than the IMU sensors, so the former is used in our evaluation as 'lower quality' sensors. This setting clearly represents the motivation of this work which is the gap between laboratory and real-life settings. It is common to see that the sensors in the laboratory are of higher quality and are controlled better than those of the final users.

\begin{figure}
    \centering
    \includegraphics[width=0.45\columnwidth]{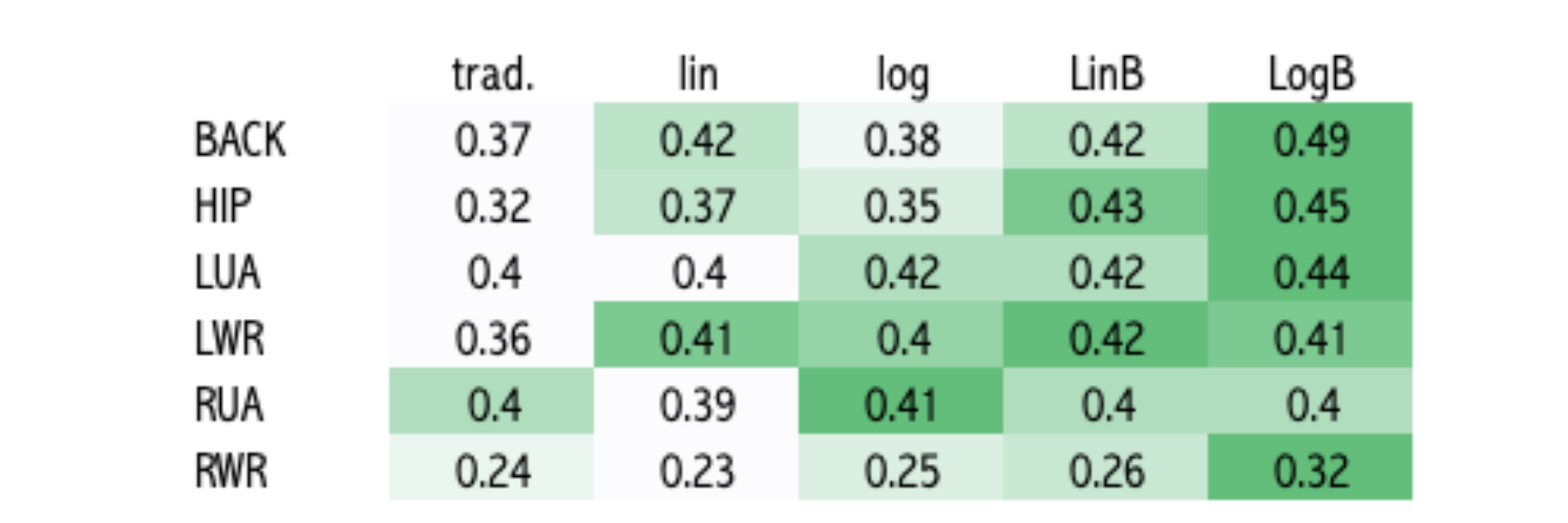}
    \caption{\textbf{F1-Score when using lower quality sensors as testing:} Results in the OPP-HL Dataset suggests that the method can be used with sensors that are of lower quality than those used in learning the representation. The improvements are as much as 12 percentage points for the HIP and BACK sensors. }
    \label{fig:low_quality}
\end{figure}

For this setting, we used the IMU sensors for learning the representation, and the lower quality acceleration sensors for learning the mapping and classification function.
Figure~\ref{fig:low_quality} shows the performance of the proposed approach in this setting. Notice that the performance of the traditional single-sensor approach is lower when high-quality sensors are used (Figure~\ref{fig:all_approaches_report}). However, the improvements are visible when using the proposed approach instead of the traditional approach. In this case, we observe  \emph{improvements of 12 percentage points for the back sensor, 13 percentage points for the HIP sensor, 8 percentage points for the RWR sensor and 5 percentage points of the LWR sensor}. The case of the hip sensor is interesting because, in the dataset used to learn the representation, there is no HIP sensor. This means that the positive transfer can occur even at new sensor positions.

This result shows that the proposed approach can be used to learn models that will use different sensors than those used in the laboratory.

\subsubsection{Using Different Types of Sensors for Feature Learning}
\label{sec:more_sensors}
We noticed that if we include different types of sensors during feature learning, the performance of the proposed approach will improve. For this, we included not only the accelerometer measures as before but also the gyroscope and magnetometer measures of the Inertial Measurement Units during the learning features step.

\begin{figure}
    \centering
    \includegraphics[width=0.95\columnwidth]{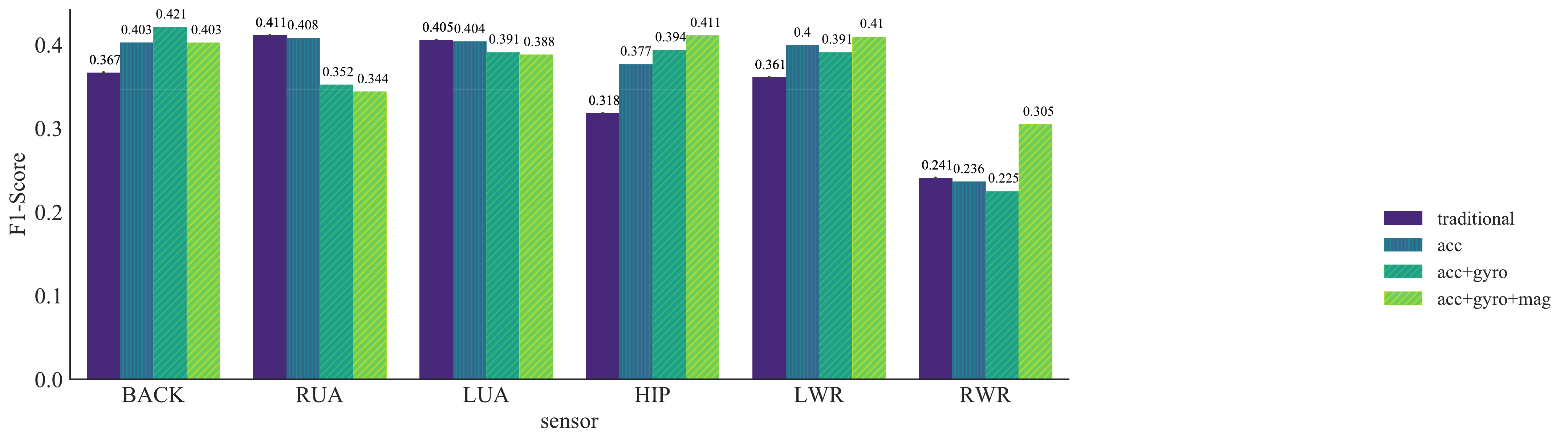}
    \caption{\textbf{F1-Score as more types of sensors are used for learning features} Including different types of sensors for learning the representation space can further increase the improvements, notably for HIP (+10pp) and RWR (+6pp) sensors. }
    \label{fig:all_sensors}
\end{figure}

For this experiment, we used the OPP-HL dataset and the low-quality sensors for testing. We did not use the boosting approach to assess the impact of including additional information when training. The F1-Scores for each accelerometer are shown in Figure~\ref{fig:all_sensors}.
As shown in the Figure, The HIP sensor shows a consistent improvement as more sensors are included for training. The improvement compared to the traditional method is almost 10 percentage points.
The BACK sensor shows an improvement when the gyroscopes are included, however, when the magnetometer is also included, the performance is equal to that of training with only the accelerometers.
For the LWR and RWR sensors, performance decreases when the gyroscope measurements are included but it increases again when the magnetometer is included.
Interestingly, for the RWR sensor, when using all three types of sensors, the proposed method outperforms the traditional method, which had only been outperformed in the previous evaluation~(Figure~\ref{fig:low_quality}). In contrast, for the RUA sensor, the performance decreases as more sensors are used; the same phenomenon is observed for the LUA sensor.

This result shows that the improvement depends on the type of information that is included for learning the representation space, but also in the testing sensor information and its placement.

\section{Discussion}
\label{sec:discussion}

This study was designed to determine whether or not we can use additional information sources to achieve activity recognition with a single sensor. Our results show that our proposed method can improve the performance of single-sensor activity recognition, notably for complex activities. In this section, we analyze these results and the implications of this study.

\subsection{Success Factors}
Our results show that using multiple sensors for learning a common representation space can improve the performance of single-sensor-based activity classification models. The main question that this research raises is: How does using additional information help in the model? For example, if the final user uses a watch how does having information from the leg help in the classification?

As we showed in the method section, a single position cannot differentiate among multiple activities. Instead, each sensor is better at discerning different pairs of activities. In the learned representation, the combined knowledge can help discern among more activities. Our intuition is that there is an underlying representation of the activity, similar to the shared representation of multiple related tasks~\cite{Argyriou2006}. Even if the dimensions of the training and testing data are different, both sets are related because they are generated by the same activities. This relation can enable the positive transfer of knowledge~\cite{transferlearningsurvey-pan2010}. The shared representation can be understood as the actions of which the activities are composed of~\cite{Lago2017, Liu2016, Chaaraoui2012}. It has been shown that manually designed action primitives, such as \textsc{cut} or \textsc{reach}, help to recognize complex activities that are composed of many such primitives~\cite{Manzoor:2013:AID:2594708.2594712}. However, designing useful primitives for each domain and obtaining labeled data for learning them can be a difficult and time-consuming task. Therefore, our proposal first learns such primitives from the multi-sensor data.


The success of our proposed approach depends on two conditions; (1) finding an effective representation and (2) minimizing the mapping error. As is the case of the PAMAP dataset, when the representation is not effective, i.e. its performance is lower than that of a single sensor, the proposed approach is not useful. However, when the representation is effective, we have seen that the improvements can reach 17 percentage points as was the case of the Cooking dataset~(Figure~\ref{fig:cooking-dataset}). More interestingly, the case of the OPP-HL when using sensors of lower quality for testing~(Figure~\ref{fig:low_quality}) demonstrated that \emph{the proposed approach can be used to improve the performance of activity recognition systems in realistic conditions by training with high-quality data}.

The mapping error depends on the test sensor position and its relation to the learned representation, as would be the case in other regression cases. We hypothesized that learning such representation is "easier" than learning to recognize complex activities. This is because each feature in the learned representation can be thought of as an "action" or a less complex activity. This is supported by our results: for the complex activities, the Cooking and OPP-HL datasets, we observe improvements, but for the simpler activities, the OPP-Loc and PAMAP dataset, we do not observe significant improvements. This shows that the proposed method is more useful in recognizing complex activities.

\subsection{Combining with the Traditional Approach}
As the mapping error is not zero, we proposed to use boosting to combine the proposed approach and the traditional activity recognition approach.
We observed that using boosting obtained the best performance and the use of boosting improved the performance compared to that of no-boosting in most cases~(Figure~\ref{fig:all_approaches_report}). This determines that boosting can reduce the effect of mapping errors.  We also observed that boosting can improve the score, even beyond that of the classification using learned features. For example, in the PAMAP dataset, the clusters classification F1-Score was 0.42 but the boosted approach for the chest sensor achieved an F1-Score of 0.47. This is because boosting uses the features of the original sensor and the shared representation space, and thus, can recognize those activities that were already well classified with the single sensor.

\subsection{Implications of the Study}
The results we have presented raise the possibility that if we have good additional information, then we can leverage this by using the proposed method and improve the performance of activity recognition. The additional information can come from additional sensors~(Section~\ref{sec:effective_complex},\ref{sec:effective_boost}), sensors of higher quality~(Section~\ref{sec:lower_quality}),  or from sensors of different types~(Section~\ref{sec:more_sensors}). These findings raise questions regarding how to find good additional information and how to choose it.
This is an important issue for future research. Our results show that having more additional sensors is better, especially placed in different body positions (upper and lower body placements).

There are two main points for future research about the potential improvements for this method: (1) evaluating different classifiers and (2) evaluating different feature learning methods. In this study, we only used SVM as a classifier to focus on the influence of the proposed approach. Based on our results, we consider that if one classifier performs better for the traditional approach, it will also perform better with the proposed approach. As the proposed method is general and can be used with any classification algorithm, further research should be undertaken to investigate the effects of using different classifiers.  Similarly, we used feature clustering as a feature learning method but as discussed in the related work, several feature learning methods could be used. Since the improvements depend on the performance of these features, using better approaches for feature learning would yield higher improvements. However, it is important to keep the dimensionality of the representation space low so that the mapping function can be learned. A further study with more focus on feature learning is suggested.

\section{Conclusions}
\label{sec:conclusions}

In this paper, we have proposed and evaluated a new activity recognition approach that takes advantage of additional information collected during experimentation or short periods in the real-life setting. Our main motivation is the observation that though it is possible to use multiple high precision sensors during short time intervals, it is not always possible to use it in real-life continuously.
Due to the fact that multi-dimensional information can better characterize an activity, our approach aims at using this knowledge to recognize different activities easier.

Our proposed approach is based on feature learning using multi-dimensional and high-precision information. We hypothesize that this new representation space encapsulates knowledge about motion movements like actions or other simple gestures. By mapping single sensor features to such representation spaces, we can effectively model activities. While our results show modest improvements, they imply that performance (as measured by the F1-score) of an accelerometer-based activity recognition model can be improved through the use of additional information for training, notably, the performance on complex and middle-level activities.

For future works, we will focus on the evaluation of the proposed approach using laboratory data as training data for real-life settings to study if the knowledge obtained in experimental settings can be successfully transferred.

The proposed approach has practical applications in fields other than activity recognition. The condition of having access to specialized sensors only for a short time can occur in other domains involving behavior studies where cameras can be used for recording a small sample.

\bibliographystyle{plain}
\bibliography{08_references}

\begin{thebibliography}{10}

\bibitem{ABBATE2012883}
Stefano Abbate, Marco Avvenuti, Francesco Bonatesta, Guglielmo Cola, Paolo
  Corsini, and Alessio Vecchio.
\newblock A smartphone-based fall detection system.
\newblock {\em Pervasive and Mobile Computing}, 8(6):883 -- 899, 2012.
\newblock Special Issue on Pervasive Healthcare.

\bibitem{Argyriou2006}
Andreas Argyriou, Theodoros Evgeniou, and Massimiliano Pontil.
\newblock Multi-task feature learning.
\newblock In {\em Proceedings of the 19th International Conference on Neural
  Information Processing Systems}, NIPS’06, page 41–48, Cambridge, MA, USA,
  2006. MIT Press.

\bibitem{Bao2004}
Ling Bao and Stephen~S. Intille.
\newblock Activity recognition from user-annotated acceleration data.
\newblock In Alois Ferscha and Friedemann Mattern, editors, {\em Pervasive
  Computing}, pages 1--17, Berlin, Heidelberg, 2004. Springer Berlin
  Heidelberg.

\bibitem{Bhattacharya2014}
Sourav Bhattacharya, Petteri Nurmi, Nils Hammerla, and Thomas Pl{\"{o}}tz.
\newblock {Using unlabeled data in a sparse-coding framework for human activity
  recognition}.
\newblock {\em Pervasive and Mobile Computing}, 15:242--262, 2014.

\bibitem{Blanke2010}
Ulf Blanke and Bernt Schiele.
\newblock {Remember and transfer what you have learned - recognizing composite
  activities based on activity spotting}.
\newblock In {\em International Symposium on Wearable Computers (ISWC) 2010},
  pages 1--8, Seoul, South Korea, oct 2010. IEEE.

\bibitem{Chaaraoui2012}
Alexandros~Andr{\'{e}} Chaaraoui, Pau Climent-P{\'{e}}rez, and Francisco
  Fl{\'{o}}rez-Revuelta.
\newblock {A review on vision techniques applied to Human Behaviour Analysis
  for Ambient-Assisted Living}.
\newblock {\em Expert Systems with Applications}, 39(12):10873--10888, 2012.

\bibitem{6208895}
L.~{Chen}, J.~{Hoey}, C.~D. {Nugent}, D.~J. {Cook}, and Z.~{Yu}.
\newblock Sensor-based activity recognition.
\newblock {\em IEEE Transactions on Systems, Man, and Cybernetics, Part C
  (Applications and Reviews)}, 42(6):790--808, Nov 2012.

\bibitem{Cheng2013}
Heng-Tze Cheng, Martin Griss, Paul Davis, Jianguo Li, and Di~You.
\newblock Towards zero-shot learning for human activity recognition using
  semantic attribute sequence model.
\newblock In {\em Proceedings of the 2013 ACM International Joint Conference on
  Pervasive and Ubiquitous Computing}, UbiComp ’13, page 355–358, New York,
  NY, USA, 2013. Association for Computing Machinery.

\bibitem{Cook2013}
Diane Cook, Kyle~D. Feuz, and Narayanan~C. Krishnan.
\newblock Transfer learning for activity recognition: a survey.
\newblock {\em Knowledge and Information Systems}, 36(3):537--556, Sep 2013.

\bibitem{Favela2013}
J.~{Favela}.
\newblock Behavior-aware computing: Applications and challenges.
\newblock {\em IEEE Pervasive Computing}, 12(3):14--17, July 2013.

\bibitem{Guan2007}
D.~{Guan}, W.~{Yuan}, Y.~{Lee}, A.~{Gavrilov}, and S.~{Lee}.
\newblock Activity recognition based on semi-supervised learning.
\newblock In {\em 13th IEEE International Conference on Embedded and Real-Time
  Computing Systems and Applications (RTCSA 2007)}, pages 469--475, Daegu, Aug
  2007. IEEE.

\bibitem{hastie2009multi}
Trevor Hastie, Saharon Rosset, Ji~Zhu, and Hui Zou.
\newblock Multi-class adaboost.
\newblock {\em Statistics and its Interface}, 2(3):349--360, 2009.

\bibitem{Hoey2012}
Jesse Hoey, Craig Boutilier, Pascal Poupart, Patrick Olivier, Andrew Monk, and
  Alex Mihailidis.
\newblock {People, sensors, decisions: Customizable and Adaptive Technologies
  for Assistance in Healthcare}.
\newblock {\em ACM Transactions on Interactive Intelligent Systems},
  2(4):1--36, 2012.

\bibitem{Hu2011}
Derek~Hao Hu, Vincent~Wenchen Zheng, and Qiang Yang.
\newblock {Cross-domain activity recognition via transfer learning}.
\newblock {\em Pervasive and Mobile Computing}, 7(3):344--358, 2011.

\bibitem{Jain2018}
Ankita Jain and Vivek Kanhangad.
\newblock {Human Activity Classification in Smartphones Using Accelerometer and
  Gyroscope Sensors}.
\newblock {\em IEEE Sensors Journal}, 18(3):1169--1177, 2018.

\bibitem{cookingdataset}
Frank Krüger, Martin Nyolt, Kristina Yordanova, Albert Hein, and Thomas
  Kirste.
\newblock Computational state space models for activity and intention
  recognition. a feasibility study.
\newblock {\em PLOS ONE}, 9(11):1--24, 11 2014.

\bibitem{Lago2017}
Paula Lago, Claudia Jim{\'{e}}nez-Guar{\'{i}}n, and Claudia Roncancio.
\newblock {Contextualized behavior patterns for change reasoning in Ambient
  Assisted Living: A formal model}.
\newblock {\em Expert Systems}, 34(2):e12163, apr 2017.

\bibitem{5206594}
C.~H. {Lampert}, H.~{Nickisch}, and S.~{Harmeling}.
\newblock Learning to detect unseen object classes by between-class attribute
  transfer.
\newblock In {\em 2009 IEEE Conference on Computer Vision and Pattern
  Recognition}, pages 951--958, Miami, FL, USA, June 2009. IEEE.

\bibitem{survey-Lara2013}
Oscar~D. Lara and Miguel~A. Labrador.
\newblock {A Survey on Human Activity Recognition using Wearable Sensors}.
\newblock {\em IEEE Communications Surveys {\&} Tutorials}, 15(3):1192--1209,
  2013.

\bibitem{tanzeem_practical}
Jonathan Lester, Tanzeem Choudhury, and Gaetano Borriello.
\newblock A practical approach to recognizing physical activities.
\newblock In Kenneth~P. Fishkin, Bernt Schiele, Paddy Nixon, and Aaron Quigley,
  editors, {\em Pervasive Computing}, pages 1--16, Berlin, Heidelberg, 2006.
  Springer Berlin Heidelberg.

\bibitem{JingenLiu:2011:RHA:2191740.2191802}
Jingen Liu, B.~Kuipers, and S.~Savarese.
\newblock Recognizing human actions by attributes.
\newblock In {\em Proceedings of the 2011 IEEE Conference on Computer Vision
  and Pattern Recognition}, CVPR '11, pages 3337--3344, Washington, DC, USA,
  2011. IEEE Computer Society.

\bibitem{Liu2016}
Ye~Liu, Liqiang Nie, Li~Liu, and David~S. Rosenblum.
\newblock {From action to activity: Sensor-based activity recognition}.
\newblock {\em Neurocomputing}, 181:108--115, mar 2016.

\bibitem{semisupervised_tanzeem:2007}
Maryam Mahdaviani and Tanzeem Choudhury.
\newblock Fast and scalable training of semi-supervised crfs with application
  to activity recognition.
\newblock In {\em Proceedings of the 20th International Conference on Neural
  Information Processing Systems}, NIPS'07, pages 977--984, USA, 2007. Curran
  Associates Inc.

\bibitem{Manzoor:2013:AID:2594708.2594712}
Atif Manzoor, Hong-Linh Truong, Alberto Calatroni, Daniel Roggen, M{\'e}lanie
  Bouroche, Siobh\'{a}n Clarke, Vinny Cahill, Gerhard Tr\"{o}ster, and Schahram
  Dustdar.
\newblock Analyzing the impact of different action primitives in designing
  high-level human activity recognition systems.
\newblock {\em J. Ambient Intell. Smart Environ.}, 5(5):443--461, September
  2013.

\bibitem{7398506}
R.~Matsushige, K.~Kakusho, and T.~Okadome.
\newblock Semi-supervised learning based activity recognition from sensor data.
\newblock In {\em 2015 IEEE 4th Global Conference on Consumer Electronics
  (GCCE)}, pages 106--107, Osaka, Japan, Oct 2015. IEEE.

\bibitem{Natarajan2016}
Annamalai Natarajan, Gustavo Angarita, Edward Gaiser, Robert Malison, Deepak
  Ganesan, and Benjamin~M. Marlin.
\newblock Domain adaptation methods for improving lab-to-field generalization
  of cocaine detection using wearable ecg.
\newblock In {\em Proceedings of the 2016 ACM International Joint Conference on
  Pervasive and Ubiquitous Computing}, UbiComp ’16, page 875–885, New York,
  NY, USA, 2016. Association for Computing Machinery.

\bibitem{Nguyen:2015:IDS:2750858.2804256}
Le~T. Nguyen, Ming Zeng, Patrick Tague, and Joy Zhang.
\newblock I did not smoke 100 cigarettes today!: Avoiding false positives in
  real-world activity recognition.
\newblock In {\em Proceedings of the 2015 ACM International Joint Conference on
  Pervasive and Ubiquitous Computing}, UbiComp '15, pages 1053--1063, New York,
  NY, USA, 2015. ACM.

\bibitem{Nguyen:2015:RNA:2802083.2808388}
Le~T. Nguyen, Ming Zeng, Patrick Tague, and Joy Zhang.
\newblock Recognizing new activities with limited training data.
\newblock In {\em Proceedings of the 2015 ACM International Symposium on
  Wearable Computers}, ISWC '15, pages 67--74, New York, NY, USA, 2015. ACM.

\bibitem{Pan2008}
Sinno~Jialin Pan, James~T. Kwok, and Qiang Yang.
\newblock {Transfer learning via dimensionality reduction}.
\newblock {\em Proceedings of the National Conference on Artificial
  Intelligence}, 2:677--682, 2008.

\bibitem{transferlearningsurvey-pan2010}
Sinno~Jialin Pan and Qiang Yang.
\newblock {A survey on transfer learning}.
\newblock {\em IEEE Transactions on Knowledge and Data Engineering},
  22(10):1345--1359, 2010.

\bibitem{Plotz2011}
Thomas Pl{\"o}tz, Nils~Y. Hammerla, and Patrick Olivier.
\newblock {Feature Learning for Activity Recognition in Ubiquitous Computing}.
\newblock In {\em Proceedings of the Twenty-Second International Joint
  Conference on Artificial Intelligence - Volume Two}, pages 1729--1734,
  Barcelona, Catalonia, Spain, 2011. AAAI Press.

\bibitem{Raina2007}
Rajat Raina, Alexis Battle, Honglak Lee, Benjamin Packer, and Andrew~Y. Ng.
\newblock {Self-taught learning}.
\newblock In {\em Proceedings of the 24th international conference on Machine
  learning - ICML '07}, pages 759--766, New York, New York, USA, 2007. ACM
  Press.

\bibitem{Rashidi2011}
Parisa Rashidi and Diane~J. Cook.
\newblock {Activity knowledge transfer in smart environments}.
\newblock {\em Pervasive and Mobile Computing}, 7(3):331--343, jun 2011.

\bibitem{PAMAPDataset}
A.~{Reiss} and D.~{Stricker}.
\newblock Introducing a new benchmarked dataset for activity monitoring.
\newblock In {\em 2012 16th International Symposium on Wearable Computers},
  pages 108--109, Newcastle, UK, June 2012. IEEE.

\bibitem{OPP_Dataset}
D.~{Roggen}, A.~{Calatroni}, M.~{Rossi}, T.~{Holleczek}, K.~{Förster},
  G.~{Tröster}, P.~{Lukowicz}, D.~{Bannach}, G.~{Pirkl}, A.~{Ferscha},
  J.~{Doppler}, C.~{Holzmann}, M.~{Kurz}, G.~{Holl}, R.~{Chavarriaga},
  H.~{Sagha}, H.~{Bayati}, M.~{Creatura}, and J.~d.~R.~{Millàn}.
\newblock Collecting complex activity datasets in highly rich networked sensor
  environments.
\newblock In {\em 2010 Seventh International Conference on Networked Sensing
  Systems (INSS)}, pages 233--240, Kassel, Germany, June 2010. IEEE.

\bibitem{adARC-Roggen2013}
Daniel Roggen, Kilian F{\"{o}}rster, Alberto Calatroni, and Gerhard
  Tr{\"{o}}ster.
\newblock {The adARC pattern analysis architecture for adaptive human activity
  recognition systems}.
\newblock {\em Journal of Ambient Intelligence and Humanized Computing},
  4(2):169--186, 2013.

\bibitem{Saeed2019}
Aaqib Saeed, Tanir Ozcelebi, and Johan Lukkien.
\newblock {Multi-task Self-Supervised Learning for Human Activity Detection}.
\newblock {\em Proceedings of the ACM on Interactive, Mobile, Wearable and
  Ubiquitous Technologies}, 3(2):1--30, 2019.

\bibitem{Samarah2018}
Samer Samarah, Mohammed GH~AL Zamil, Majdi Rawashdeh, M.~Shamim Hossain, Ghulam
  Muhammad, and Atif Alamri.
\newblock {Transferring activity recognition models in FOG computing
  architecture}.
\newblock {\em Journal of Parallel and Distributed Computing}, 122:122--130,
  dec 2018.

\bibitem{Shirahama2016}
Kimiaki Shirahama, Lukas K\"{o}ping, and Marcin Grzegorzek.
\newblock Codebook approach for sensor-based human activity recognition.
\newblock In {\em Proceedings of the 2016 ACM International Joint Conference on
  Pervasive and Ubiquitous Computing: Adjunct}, UbiComp ’16, page 197–200,
  New York, NY, USA, 2016. Association for Computing Machinery.

\bibitem{Stikic2011}
M.~{Stikic}, D.~{Larlus}, S.~{Ebert}, and B.~{Schiele}.
\newblock Weakly supervised recognition of daily life activities with wearable
  sensors.
\newblock {\em IEEE Transactions on Pattern Analysis and Machine Intelligence},
  33(12):2521--2537, Dec 2011.

\bibitem{5254653}
M.~Stikic, D.~Larlus, and B.~Schiele.
\newblock Multi-graph based semi-supervised learning for activity recognition.
\newblock In {\em 2009 International Symposium on Wearable Computers}, pages
  85--92, Linz, Austria, Sept 2009. IEEE.

\bibitem{VanKasteren2010}
T.~L.~M. van Kasteren, G.~Englebienne, and B.~J.~A. Kr\"{o}se.
\newblock Transferring knowledge of activity recognition across sensor
  networks.
\newblock In {\em Proceedings of the 8th International Conference on Pervasive
  Computing}, Pervasive'10, pages 283--300, Berlin, Heidelberg, 2010.
  Springer-Verlag.

\bibitem{Vapnik2015}
Vladimir Vapnik.
\newblock {Learning Using Privileged Information : Similarity Control and
  Knowledge Transfer}.
\newblock {\em Journal of Machine Learning Research}, 16:2023--2049, 2015.

\bibitem{WEN201719}
Jiahui Wen and Zhiying Wang.
\newblock Learning general model for activity recognition with limited labelled
  data.
\newblock {\em Expert Systems with Applications}, 74:19 -- 28, 2017.

\bibitem{Xu2016}
Tiantian Xu, Fan Zhu, Edward~K. Wong, and Yi~Fang.
\newblock {Dual many-to-one-encoder-based transfer learning for cross-dataset
  human action recognition}.
\newblock {\em Image and Vision Computing}, 55:127--137, 2016.

\bibitem{xu16dsne}
X.~{Xu}, X.~{Zhou}, R.~{Venkatesan}, G.~{Swaminathan}, and O.~{Majumder}.
\newblock d-sne: Domain adaptation using stochastic neighborhood embedding.
\newblock In {\em 2019 IEEE/CVF Conference on Computer Vision and Pattern
  Recognition (CVPR)}, pages 2492--2501, Long Beach, CA, USA, June 2019. IEEE.

\bibitem{yang1999evaluation}
Yiming Yang.
\newblock An evaluation of statistical approaches to text categorization.
\newblock {\em Information retrieval}, 1(1-2):69--90, 1999.

\end{thebibliography}

\end{document}